\documentclass[prd,aps,graphics]{revtex4}
\usepackage{epsfig}

\begin{document}

%\magnification\magstep1
%\openup 1\jot
\font\cst=cmr10 scaled \magstep3
\font\csc=cmr10 scaled \magstep2
\vglue 2.5cm

\centerline{\cst  Bouncing universes and their perturbations~:}
\vskip 0.5cm
 \centerline{\cst remarks on a toy model}

\vskip 1 true cm
\centerline{\bf Nathalie Deruelle}
\vskip 0.5cm
\centerline{\it Institut d'Astrophysique de Paris,}
\centerline{\it GReCO, FRE 2435 du CNRS,}
\centerline{\it 98 bis boulevard  Arago, 75014, Paris, France}

\centerline{and}

\centerline{\it Institut des Hautes Etudes Scientifiques,}
\centerline{\it 35 Route de Chartres, 91440, Bures-sur-Yvette, France}

\medskip
\vskip 1cm
\centerline{27 April 2004}
\centerline{\it Version 2 (13/06/04). Expands a number of points for clarity. Conclusions unchanged.}

\vskip 3.5cm
\noindent
{\bf Abstract}
\bigskip

Friedmann-Lema\^\i tre universes driven by a scalar field, spatially closed and bouncing, were recently studied in [1], with the conclusion that the spectrum of their large scale
matter perturbations was generically modified when going through the bounce. In this Note we extend this result to a wider class of bouncing scale factors and give the
properties of the scalar field potentials which drive them. In doing so we throw light on the hypotheses which underlie the models and discuss their cogency.

\vfill\eject

\noindent
{\csc I. Introduction}

\vskip 0.2cm

Since the invention of the string-inspired, ``pre-Big Bang" [2] and ``ekpyrotic" or ``cyclic" [3] universes, there has been  renewed interest in 4-dimensional, general relativistic,
bouncing, Friedmann-Lema\^\i tre models, to which they could reduce within some effective theory limit. Despite many attempts however (see e.g. [4]) an issue still under debate
is how the spectrum of initial, pre-``Big Bang", matter fluctuations is affected by the bounce.

To try and answer that question, a generic model for bouncing universes was studied in [1] (see also [5])~:   spatially closed Friedmann-Lema\^\i tre models, driven by a
scalar field, whose scale factors near the bounce are expanded as a Taylor series in conformal time with a priori arbitrary coefficients. The conclusion of that study was that
the spectrum of large scale matter perturbations was generically modified when going through the bounce. 

Two hypotheses underlie that conclusion. One is that the scalar field conformal gradient remains small at the bounce. The other provides the (required) additional
information on the asymptotic behaviour of the scale factors (not encoded in a low order Taylor series) and is in practice, as we shall see, to  approximate the scale
factor in the bouncing region by a truncated Taylor expansion, that is by a polynomial in conformal time.

 In this Note, we first show in Section 2  that approximating the scale factor in the bouncing region by a
polynomial amounts, when the scalar field conformal gradient is small at the bounce, to more than expanding it in a Taylor series. Indeed, as we shall see, the crucial fact that
the effective potential for the perturbations decays at the outskirts of the bouncing region (and hence allows to define the transition matrix for the spectrum) is an information
which cannot be obtained by a low order Taylor expansion but stems from the fact that the scale factor is approximated by a polynomial. We will however extend the
class of scale factors which yield decaying  potentials for the perturbations to other generic functions of time,  which are not truncated to low order polynomials in the bouncing
region.

We then bridge a gap left open in [1] and turn in Section 3 to the characteristics of the scalar potentials required to reach the conclusion that the perturbation spectrum is
modified as described in [1]. We find they are sharply peaked at the bounce and decay at its outskirts. We then study various simple scalar potentials exhibiting the same
dependance in the scalar field at the bounce and which decay at its outskirts, and find that they also lead to modifications of the perturbation spectrum, but that they must be
fairly fine tuned in the transition range to avoid premature recollapse of the universe.

Finally we discuss in Section 4 the hypothesis that the scalar field conformal gradient remains small at the bounce and argue that it can be relaxed. If, then, the scalar
field conformal gradient is allowed to be of order unity at the bounce, then, as we shall see, either the spectrum is not modified or the universe recollapses
immediately---unless the scalar field potential remains sharply peaked.

Section 5 summarizes our conclusions.

\bigskip

\noindent
{\csc II. Bouncing scale factors~: Taylor expansions {\it vs} asymptotic behaviours}

\vskip 0.2 cm
This Section summarizes [1] (see also [5]), putting the spot light on the hypothesis which is made to obtain the desired asymptotic behaviour of the perturbation 
potential, to wit,  the scale factors can be approximated by polynomials in conformal time in the bouncing region. We then extend the class of scale factors yielding similar
perturbation potentials to other generic functions of time, which are not truncated to low order polynomials in the bouncing
region.

Consider a spatially closed, homogeneous and isotropic universe with line element~: $ds^2=a^2(\eta)(-d\eta^2+d\Omega^2_3)$ where $\eta$ is the (dimensionless) conformal
time,
$a(\eta)$ the
 scale factor (with dimension of a length) and $d\Omega^2_3$ the line element of a 3-dimensional unit sphere. If matter is just a scalar field $\varphi$ with
potential
$V(\varphi)$ the Einstein equations reduce to the Friedmann-Lema\^\i tre equation~: $3({\cal H}^2+1)=\kappa\left({1\over2}\varphi'^2+a^2V\right)$ ($\kappa$ is
Einstein's constant), plus either the Klein-Gordon equation or, equivalently (unless
$\varphi'\equiv0$)~:
$${1\over2}\varphi'^2={\cal H}^2-{\cal H}'+1\eqno(2.1)$$
 where a prime
denotes differentiation with respect to conformal time and where ${\cal H}\equiv a'/a$ is the (dimensionless) Hubble parameter. In (2.1) and in the following we
set Einstein's constant equal to one (so that $\varphi$ and $\varphi'$ become dimensionless). (We also set the velocity of light to one, so that the unit of length, say, is still
unspecified).

Consider then the perturbed metric~: $ds^2=a^2(\eta)[-(1+2\Phi)d\eta^2+(1-2\Psi)d\Omega^2_3]$ and the perturbed scalar field $\varphi(\eta)+\delta\varphi$. In Fourier space,
the scalar perturbations $\Phi_n$, $\Psi_n$ and $\delta\varphi_n$ are functions of time and  of the Eigenvalues $n$ of the Laplacian on the $3$-sphere (defined as $\triangle
f_n=-n(n+2)f_n, n\in N$ and $n\geq2$). The linearized  Einstein equations  (see, e.g. [6]) then yield  two constraints $\left(\Phi_n=\Psi_n,
\delta\varphi_n={2({\cal H}\Phi_n+\Phi^{\prime}_n)\over\varphi^{\prime}}\right)$, and the following evolution equation for $u_n\equiv{a\over\varphi^\prime}\Phi_n$~:
 $$ u^{\prime\prime}_n+\left[k^2-W(\eta)\right]u_n=0\quad\hbox{where}\quad W(\eta)\equiv f+\left({f'\over2f}\right)^2-\left({f'\over2f}\right)' \quad\hbox{with}\quad
f\equiv{1\over2}\varphi'^2\eqno(2.2)$$  
and  $k^2\equiv n(n+2)-3$. According to [1], the values of $k$ which correspond to scales of cosmological interest today lie in the range
$[60-6\times 10^{6}]$.

Consider now the family of scale factors indexed by the integer $N$~:
$$a(\eta)=\sum _{n=0}^N a_{2n}\,\left({\eta\over\eta_0}\right)^{2n}\quad\hbox{with}\quad  a_0=1\ , \ a_2={1\over2}\ ,\ a_4\equiv {5(1+\xi)\over24}\ ,\
a_6\equiv{61(1+\chi)\over720}\,.\eqno(2.3)$$
As in [1] we restrict our attention to time-symmetric bounces. The value of the scale factor at the bounce is rescaled to be one (this choice sets the unit of length to be the
radius of the universe at the bounce, see Note [7]). The introduction of the parameter $\eta_0$ allows to choose the value of $a_2$ at will. $\xi$ and $\chi$ are parameters, taken
to be of order one,  used in [1].

Near the bounce, (2.3) is the  Taylor expansion to order ${\cal O}(\eta^{2N})$ of {\it any} bouncing scale factor, and yields~:
$${1\over2}\varphi'^2=\Upsilon -{5\xi\over2}\, {\eta^2\over\eta_0^4}+\left({115\over24}\xi-{61\over24}\chi\right)\,{\eta^4\over\eta_0^6}+{\cal
O}(\eta^6)\quad\hbox{where}\quad \Upsilon\equiv 1-{1\over\eta_0^2}\eqno(2.4)$$
(higher order expansions will not be necessary). Clearly, $\eta_0$ must be greater or equal to one. As in [1] we shall not consider the case $\eta_0=1$ (for which the
perturbation equation (2.2) is singular at the bounce, see Note [8]). Moreover we shall concentrate, as in [1], on the case when the (dimensionless) scalar field conformal gradient
$\varphi'$ remains small at the bounce, that is to $\Upsilon\ll1$ (we shall discuss this hypothesis in Section 4).

As for the Taylor expansion of the effective potential for the perturbations $W(\eta)$ it reads~:
$$W(\eta)=\left(\Upsilon+{5\xi\over2\eta_0^4}{1\over\Upsilon}\right)+\left[{61\chi\over4\eta_0^2\Upsilon}-{5\xi\over2}\left(1+{23\over2\eta_0^2\Upsilon}
-{10\xi\over\eta_0^4\Upsilon^2}\right)\right]{\eta^2\over\eta_0^4}+{\cal O}(\eta^4)\,.\eqno(2.5)$$
With the additional assumption $\Upsilon\ll1$, and for generic coefficients $(\xi, \chi,...)$ of order one, it no longer depends on $\chi$ and becomes, see [1]~:
$$W(\eta)\approx{5\xi\over2\Upsilon}\left(1+{10\xi\over\Upsilon}\eta^2\right)+{\cal O}(\eta^4)\quad\hbox{when}\quad \Upsilon\to0_+\,.\eqno(2.6)$$

\bigskip
Now, and this is one of the main points of this Note, in order to relate the pre-bounce perturbation modes to the post-bounce ones, one needs asymptotic regions where to define
the ``in" and ``out" states. Such asymptotic behaviours cannot in general stem from a low order Taylor expansion,
whose range of validity is, usually, limited~: the Taylor expansion of $W(\eta)$ always blows up, whatever the order in ~$\eta$. If, moreover,  $\Upsilon\ll1$ it generically
(that is for $(\xi,\chi,...)$ of order one) blows up for
$|\eta|\gg\sqrt\Upsilon$, that is, well inside the bouncing region, see (2.6). {\it Hence the assumption, made in [1] (see, e.g., equations (30) or (41) therein), to approximate the
scale factor in the bouncing region by the polynomial  (2.3)}. That extra assumption yields the following asymptotic behaviours~:
$${1\over2}\varphi'^2\approx W(\eta)=1+{2N(N+1)\over\eta^2}+{\cal O}(\eta^{-4})\eqno(2.7)$$
which, in practice (see Figure 1), are valid at the outskirts of the bouncing region, that is for $\sqrt\Upsilon\ll |\eta|\leq1$, see Note [9].

Concentrating mainly on the case $-1<\xi<0$, $N=2$, the authors of [1] were then able to approximate $W(\eta)$ by a simple rational function and  show that its shape was fairly
independent of $\xi$ and $\eta_0$, and hence generic.

Therefore, with the two hypotheses $\Upsilon\ll1$ (that is small scalar field conformal gradient at the bounce) and polynomial scale factor in the bouncing region,
$W(\eta)$ is a deep potential well (see Figure 1), which can in effect be approximated by (2.6) for
$|\eta|<\sqrt{-\Upsilon/10\xi}$, and, from (2.7), by
$W=0$ for
$|\eta|>\sqrt{-\Upsilon/10\xi}$, or, even more simply by a square potential of depth $V_m\equiv -5\xi/2\Upsilon$ and width
$\sqrt{-2\Upsilon/5\xi}$ (or, better, twice that value, see [1]).

 Today large scales correspond to modes $k^2$ small compared to the depth of the potential, if $\Upsilon$ is sufficiently small (say, $\Upsilon<10^{-6}$). It is then a simple
exercise to see that  their spectrum  is affected by the bounce. More precisely~: the pre-bounce mode, which behaves as
$u_n(\eta)=A^{\rm pre}_n\exp{(ik\eta)}+B^{\rm pre}_n\exp{(-ik\eta)}$ in the asymptotic region $1\leq\eta\ll-\sqrt{-2\Upsilon/5\xi}$ where $W(\eta)\approx0$ is the square
potential approximation of (2.7), turns into the post-bounce mode
$u_n(\eta)=A^{\rm post}_n\exp{(ik\eta)}+B^{\rm post}_n\exp{(-ik\eta)}$ for $1\geq\eta\gg\sqrt{-2\Upsilon/5\xi}$ with,  (see [1])~:

$$\left(\matrix{A_n^{\rm post}\cr B_n^{\rm post}}\right) ={\rm i} \, {Const.\over k}\sqrt{-\xi\over\Upsilon}\left(\matrix{\ \ 1&\ \ 1\ \cr
-1&-1\cr}\right)\left(\matrix{A_n^{\rm pre}\cr B_n^{\rm pre}}\right) +{\cal O}\left(\left({k\over \sqrt{V_m}}\right)^0\right)\eqno(2.8)$$
where the precise value of the $Const.(\in {\cal R})$ depends on the level of approximation made for $W(\eta)$ in the region $|\eta|\leq\sqrt{-2\Upsilon/5\xi}$. (The fact that
the transition matrix is not invertible is not surprising as we are in the limit where an incident wave is totally reflected by the deep well, see [5] for developments). The
important point is the dependance on $k$, that we
 checked by direct numerical integration, see Figure 1,  [10].

\begin{figure}
\centerline{\epsfig{figure=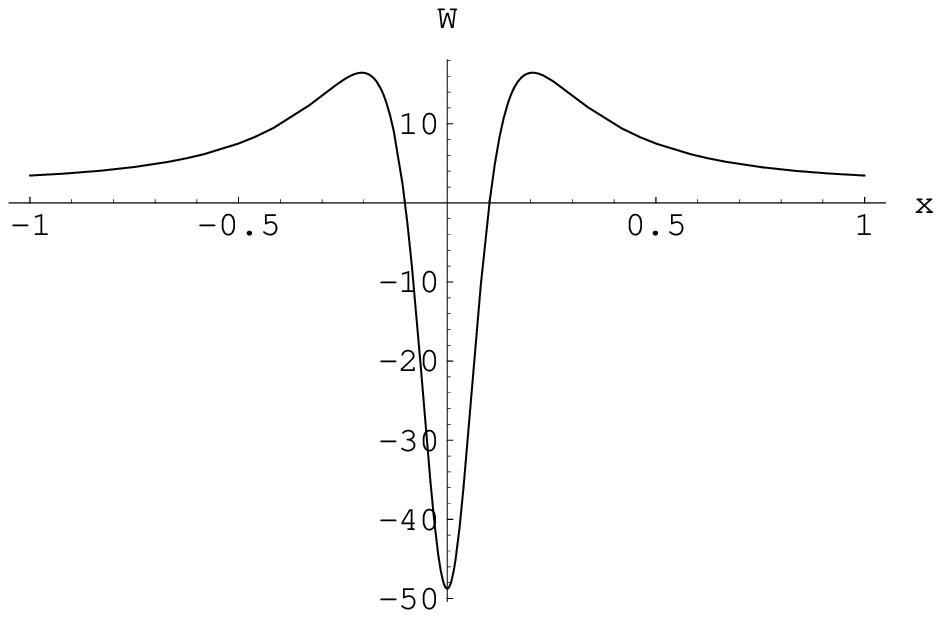,width=8cm}, \epsfig{figure=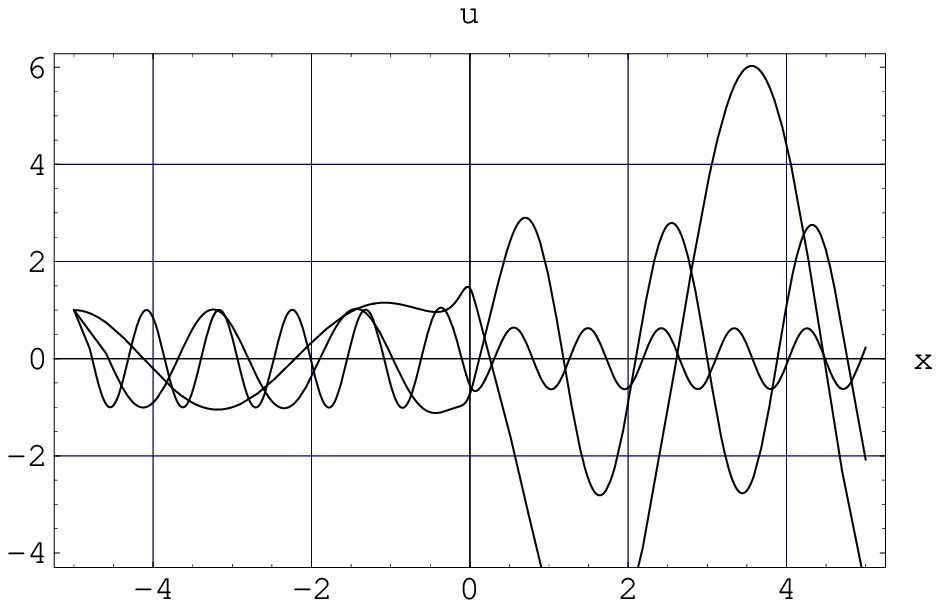,width=8cm}}
\caption{{\it Matter perturbations induced by polynomial scale factors.}  The left panel (first obtained in [1]) shows a typical deep potential well for the
perturbations $W(x)$ with $x\equiv\eta/\eta_0$, generated by a polynomial scale factor. The parameters are the same as those illustrating [1], that is~: $\eta_0=1.01$,
$\xi=-2/5$, $N=2$, see equation (2.3) for definitions. The right panel exhibits a few perturbations modes $u_n(x)$ (generated by the potential $W(x)$ on the left panel), and shows that their spectrum is modified by the bounce, according to (2.8) (see [1]) when $n^2$ is small compared to depth of $W(x)$.}
\label{Figure 1}
\end{figure}

 Other cases than $-1<\xi<0$, $N=2$ can be considered. For example, we looked at the cases~: $N=1$ (parabolic scale factor), $N=3$, $-1<\xi<0$ with $\chi>0$ or $-1<\chi<0$~:
in all those examples, the shape of the potential $W(\eta)$ is almost independant of the higher order coefficients $a_{2n}$ and of the value of $\Upsilon$ (as long as it is small),
and we checked that the result (2.8) obtained in [1] holds (at least approximately). In fact, we found that it also holds when, say,  $N=2$, $-2<\xi<-1$, that is when the
universe recollapses, because, for
$\Upsilon$ small enough, there is a region around the bounce where, first, the potential $W(\eta)$ can be approximated as above, and, second, which is large enough to define the
asymptotic states of the perturbations $u_n$. 
We also checked that scale factors polynomial in {\it cosmic} time $t$ (and, hence, non polynomial in conformal time  $d\eta=dt/a(t)$) yield the same result, that is (2.8).
Finally, another possibility is to impose that the scale factor asymptotes its quasi-de Sitter value in the range $\sqrt\Upsilon\ll|\eta|\ll1$, for example~:
$a(\eta)={1/\cos \eta}+(b_4\, \eta^4+b_6\,\eta^6...)/\cosh \eta$. As can easily be checked, such a truncation generically has the same effect as
choosing a polynomial scale factor~: it yields a potential for the perturbations similar to Figure 1, and hence, once again,  the result (2.8) obtained in [1] holds. 

Therefore the conclusion reached in [1], that is the modification of the perturbation spectrum through a bounce due to a deep potential well is quite robust, in the sense that
it holds for a large class of scale factors. As we have stressed, the condition of validity of the conclusion are that, (1), the scalar field conformal gradient remains small at
the bounce ($\Upsilon\ll1$) and, (2), that the effective potential for the perturbations, $W(\eta)$, decays for $|\eta|\gg\sqrt\Upsilon$, so that asymptotic regions for the in and
out perturbation states can be defined.

However, before claiming convincingly that all the examples studied ``disprove {\sl a priori} any general argument stating that the spectrum should propagate through the bounce
without being modified" [1], the pertinence of the hypotheses underlying the claim must be probed and, hence,  the characteristics of the potential $V(\varphi)$ of the
scalar field which drives these  bouncing  universes must be studied.

\bigskip
\noindent
{\csc III. The driving potentials}
\vskip 0.5cm

The Friedmann-Lema\^\i tre equation (which has not been used up to now) can be written as, using (2.1)~:
$$V(\eta)={1\over a^2}({\cal H}'+2{\cal H}^2+2)\,.\eqno(3.1)$$

Near the bounce, the Taylor expansion of $V(\eta)$, for any scale factor, follows from (2.3) and is, (Note [11] recalls the units chosen)~:
$$V(\eta)=(3-\Upsilon)-2\Upsilon\left(1-{5\xi\over4\eta_0^2\Upsilon}\right){\eta^2\over\eta_0^2}+\left({2\Upsilon\over3}-{5\xi\over6}-
{65\xi\over24\eta_0^2}+ {61\chi\over24\eta_0^2}\right){\eta^4\over\eta_0^4}+{\cal O}(\eta^6)\,.\eqno(3.2)$$
Since $\varphi(\eta)$ can be extracted from (2.4) by a simple integration, $V(\varphi)$ is
known. Near the bounce  its Taylor expansion is (choosing without loss of generality $\varphi(0)=0$ and $\varphi'>0$, but excluding the value $\Upsilon=0$)~:
$$V(\varphi)=(3-\Upsilon)-\left(1-{5\xi\over 4\eta_0^2\Upsilon}\right){\varphi^2\over\eta_0^2}+\left({2\Upsilon\over3}+{61\chi\over24\eta_0^2}-{5\xi\over6}-
{35\xi\over8\eta_0^2}+{25\xi\over12\eta_0^4\Upsilon}\right){\varphi^4\over4\eta_0^4\Upsilon^2}+{\cal O}(\varphi^6)\,.\eqno(3.3)$$

When, as in [1], the scale factor is approximated by a polynomial, the large $\varphi$ behaviour of $V(\varphi)$ is~: $V(\varphi)\sim 2a^{-2}$, from (3.1)~; hence
$V(\varphi)\propto\eta^{-4N}$, from (2.3), and, since $\varphi'\sim \sqrt2$ from (2.6)~:
$$V(\varphi)={2^{(2N+3)}N^2\eta_0^{4N}\over a_{2N}^2}\,{1\over\varphi^{4N}}+{\cal O}\left({1\over\varphi^{4N+2}}\right)\,.\eqno(3.4)$$
Figure 2 gives the typical shape of $V(\varphi)$ when $\Upsilon\ll1$.

\begin{figure}
\centerline{\epsfig{figure=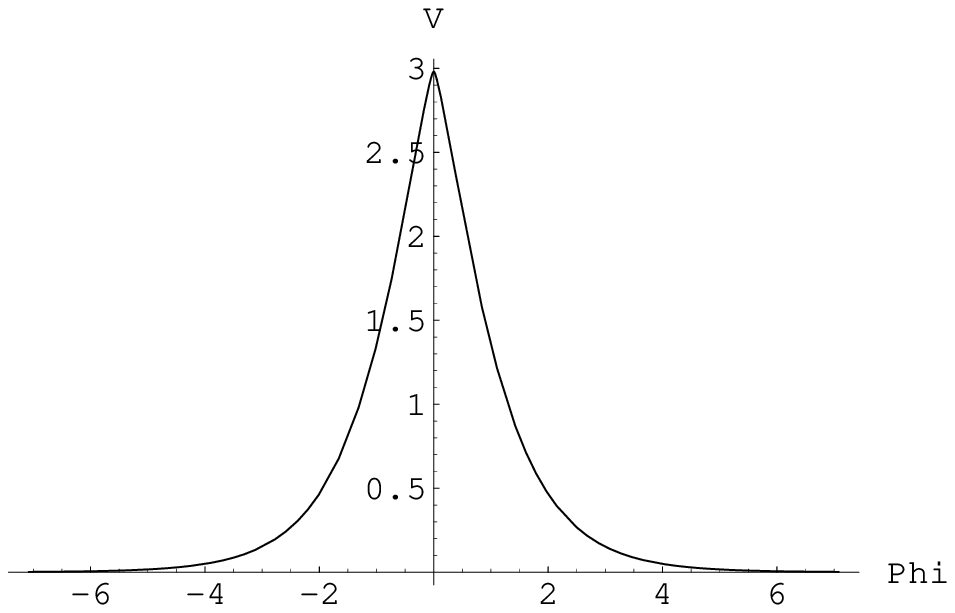,width=8cm},\epsfig{figure=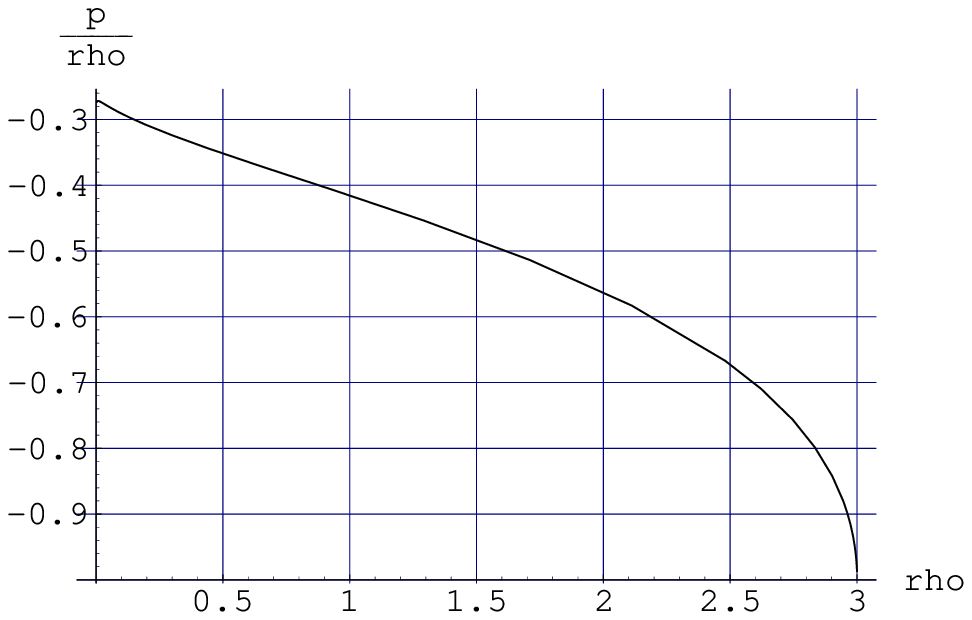,width=8cm} }
\caption{{\it Polynomial scale factors~: the driving potential.}  The left panel shows a typical driving potential $V(\varphi)$ which yields a polynomial
bouncing scale factor (the parameters are the same as those illustrating [1], see Figure 1)~; in the intermediate range, $(0<\eta<1)$, $V(\varphi)$
can be approximated by the straight line~: $V_a(\varphi)=3(1-0.56\,\varphi)$. The right panel gives the corresponding scalar field effective equation of state.}
\label{Figure 2}
\end{figure}

The main properties of $V(\varphi)$ when $\Upsilon\ll1$ are~:  (1), $V(0)\approx3$ (that is $V(0)\approx3/a_0^2$ in Planck units, see Notes [7] and [11])~;
(2), it is sharply peaked at $\varphi=0$ (as  is clear from (3.3)), that is it
exhibits an effective square mass $m^2\approx {5\xi\over2\Upsilon}$ which is negative and large (with respect to the scale set by the radius of the universe at the bounce~; in
Planck units,  $m^2\approx {5\xi\over2\Upsilon} {1\over a_0^2}$, and may be small if $a_0$ is large enough, see Note [7])~; (3), more striking perhaps, it turns out that it
decreases almost linearly in the transition region before reaching its large
$\varphi$ behaviour (3.4). Another way to characterize the underlying physics is to describe the scalar field as a fluid of density $\varrho\equiv \left({\varphi'^2\over2
a^2}+V(\varphi)\right)$ and pressure
$p\equiv\left( {\varphi'^2\over2 a^2}-V(\varphi)\right)$ and give its adiabatic index  $p/\varrho$ as a function of $\varrho$, see Figure 2. and Note [12]

Let us now turn to simple scalar potentials which resemble these ``conical hat" potentials, for example,  Lorentzians $V_{v,c}(\varphi)={v\over 1+c^2\varphi^2}$
or Gaussians
$V_{v,c}(\varphi)=v\,\rm{e}^{-c^2\varphi^2}$,  with $v$ of order one and $c$ large (we do not ask here if pre-Big-Bang or cyclic scenarios can provide us with such potentials).
It is then an easy task to see that, when one imposes $\varphi'(0)\equiv \sqrt{2\Upsilon}\ll1$ (or, equivalently, $v\approx3$), one gets  potentials
$W_{v,c}(\eta)$ for the perturbations which also exhibit a deep well and an aymptotic region, and hence for which one can also conclude that the spectrum of perturbations is
modified by the bounce, see Figure 3. Of course, the scale factors are then no longer polynomial in time. In fact, at least in the examples
considered, the universe soon recollapses (``soon" meaning for $|\eta|\approx1$), because $V_{v,c}(\varphi)$ does not decrease slowly enough in the transition region. If,
therefore,  one demands a model which not only  yields a potential $W(\eta)$  for the perturbations that exhibits a deep potential well together with asymptotic regions, but
which also avoids premature recollapse, then it seems that the potential for the scalar field must be fairly fine tuned.

\begin{figure}
\centerline{\epsfig{figure=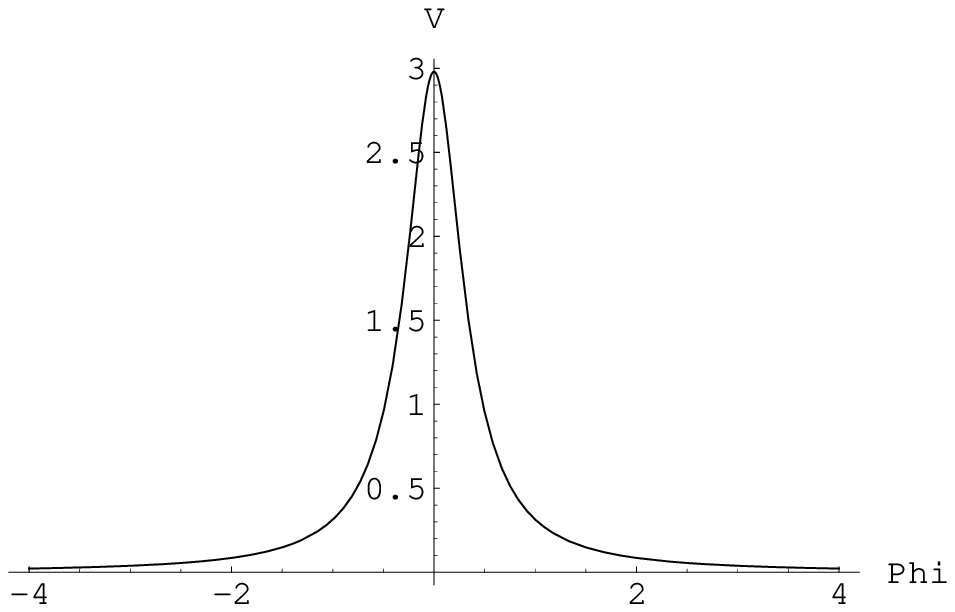,width=5.5cm},\epsfig{figure=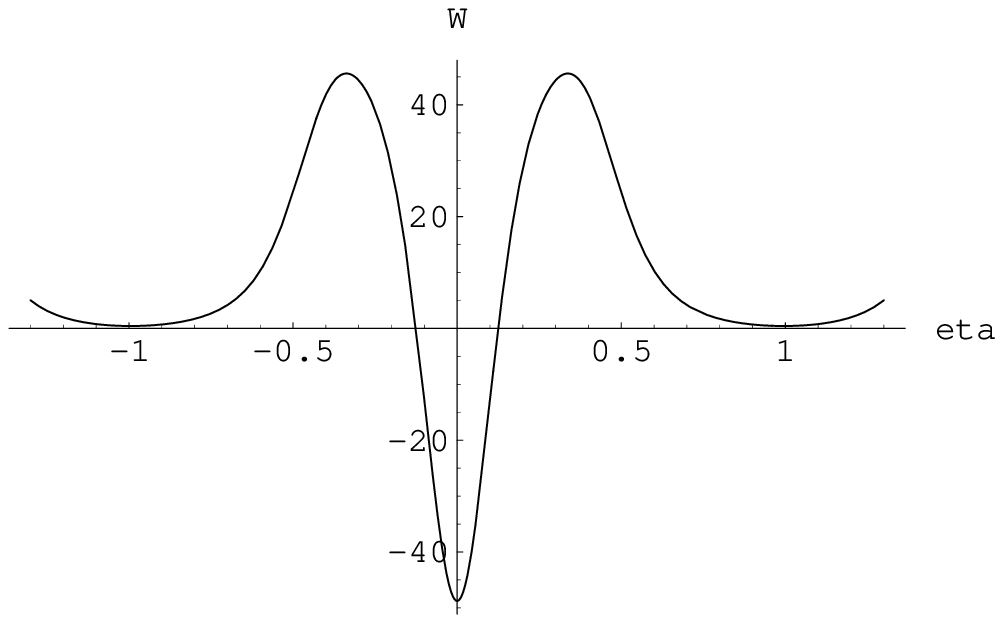,width=5.5cm}, \epsfig{figure=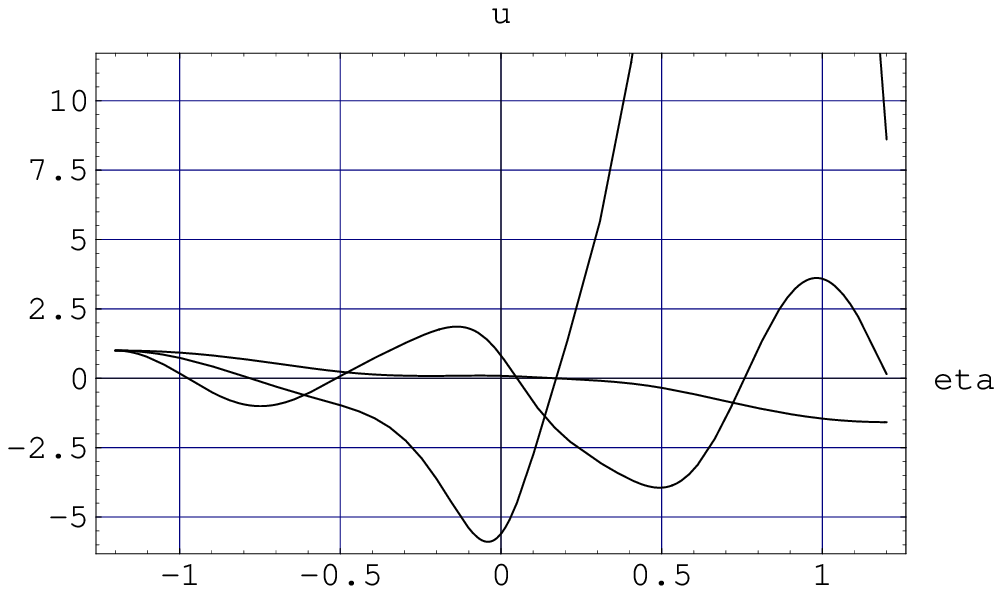,width=5.5cm}}
\caption{{\it Lorentzian scalar potentials (1).}  The left panel shows the scalar potential $V(\varphi)={v\over 1+c^2\varphi^2}$ with $v$ and $c$ such that $V(\varphi)$ has
the same small $\varphi$ behaviour as in Figure 2. The central panel shows the corresponding potential for the matter perturbations, $W(\eta)$, which must be compared to
Figure 1. (For
$\eta$ large, $W(\eta)$ diverges because the scale factor recollapses.) Finally, the right panel exhibits a few perturbations modes $u_n(x)$, with $n^2$ small compared to
the depth of $W(\eta)$, and shows that their spectrum is modified by the bounce.}
\label{Figure 3}
\end{figure}

\bigskip
\noindent
{\csc IV. Should} $\Upsilon$  {\csc be small ?}
\vskip 0.5cm

In order to avoid misunderstandings, we reestablish here the constant $a_0$ that we had previoulsy set to 1 and  use Planck units. We also suppose $a_0\gg1$ so that
the radius of the universe at the bounce is large compared to the Planck length.

In the preceding Sections we followed [1] and imposed the dimensionless scalar field conformal gradient $\varphi'(0)=\sqrt{2\Upsilon}$, see (2.4), to be small. It
seems to us that a more meaningful quantity is the {\it kinetic} energy density $E_k\equiv{1\over2}{\varphi'^2\over a^2}$ of the scalar field. At the bounce
$E_k|_0=\Upsilon/a_0^2$. Now, it turns out that, in all the cases considered (with $\Upsilon\ll1$, see end of Section 2) $E_k$ (which is even in $\eta$) increases sharply with
$\eta$ and reaches a value of order $1/a_0^2$ before gently decaying. Hence imposing $E_k|_0$ to be much smaller than its generic value in the whole bouncing region seems
somewhat unatural. 

Now, an even more meaningful quantity seems to be the {\it total} energy density of the scalar field. From Friedmann-Lema\^\i tre's equation, it is~:
$$\varrho(\eta)\equiv{1\over2}{\varphi'^2\over a^2}+V={3\over a^2}({\cal H}^2+1)\,.\eqno(4.1)$$
Its value at the bounce is $\varrho(0)=3/a_0^2$. For $a_0\gg1$, see Note [7],
$\varrho(0)$ is small compared to the Planck density, whatever the value of $\Upsilon$. Moreover,  in all cases considered,  $\varrho$ gently decreases in
${\cal O}(\eta^{-{2N}})$ in the asymptotic region, whatever the value of $\Upsilon$. Therefore, if one argues that what matters is the value of the {\it total} energy of the field
with respect to the Planck energy in the bouncing region, then $\Upsilon$ need not be small. 

Now, for $\Upsilon$ ({\it and}  $\xi$) of order one, the scalar field potential $V(\varphi)$ is not sharply
peaked at the bounce-- see (3.3)--, the potential $W(\eta)$ for the perturbations is no longer deep compared to $k^2$--see (2.5)-- and then, if there is an asymptotic region
large enough to define the in and out states, the spectrum of the perturbations is unaffected by the bounce, as stated in [1]. If, on the other hand, no asymptotic region can be
reached before recollapse (as is the case for  ``bell"-like potentials described by Lorentzians or Gaussians with both parameters of order one) then  no definite conclusion can be
drawn about the modification of the perturbation spectrum.

A last possibility must be considered though~: $\Upsilon$ of order one, {\it but} large (negative) $\xi$. This is a case when developping the scale factor in a Taylor series is
not a good strategy. Such values of the parameters can however be achieved by simple  sharply peaked potentials described by, say,  Lorentzians $V_{v,c}(\varphi)={v\over
1+c^2\varphi^2}$ with $2<v<3$, {\it and} $c$ large. As can be easily checked the potential for the perturbations and the perturbations modes are then similar to those presented
in Figure 3, the caveat being again that the decay of $V(\varphi)$ must be slow enough to prevent premature recollapse, see Figure 4.

\begin{figure}
\centerline{\epsfig{figure=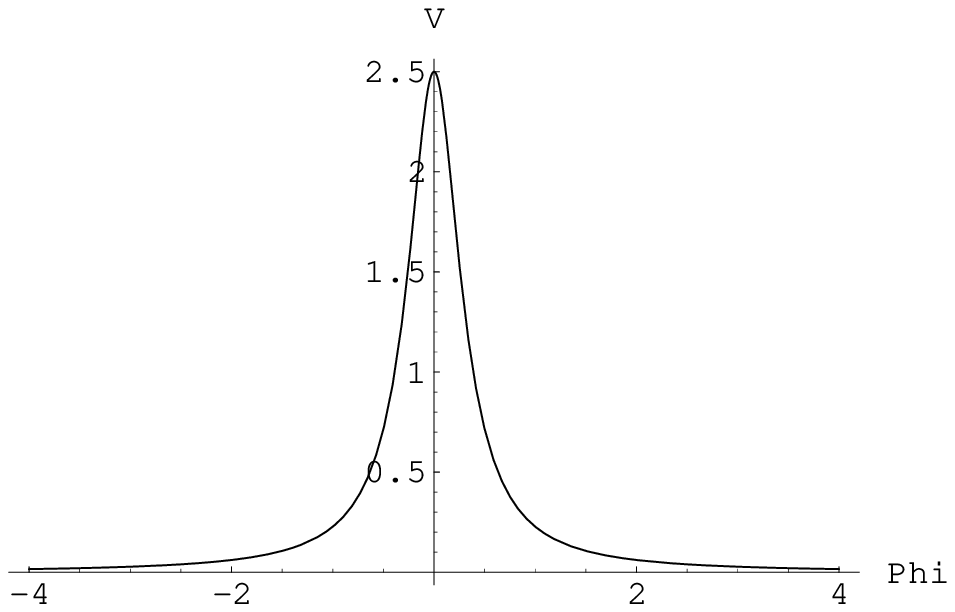,width=4cm},
\epsfig{figure=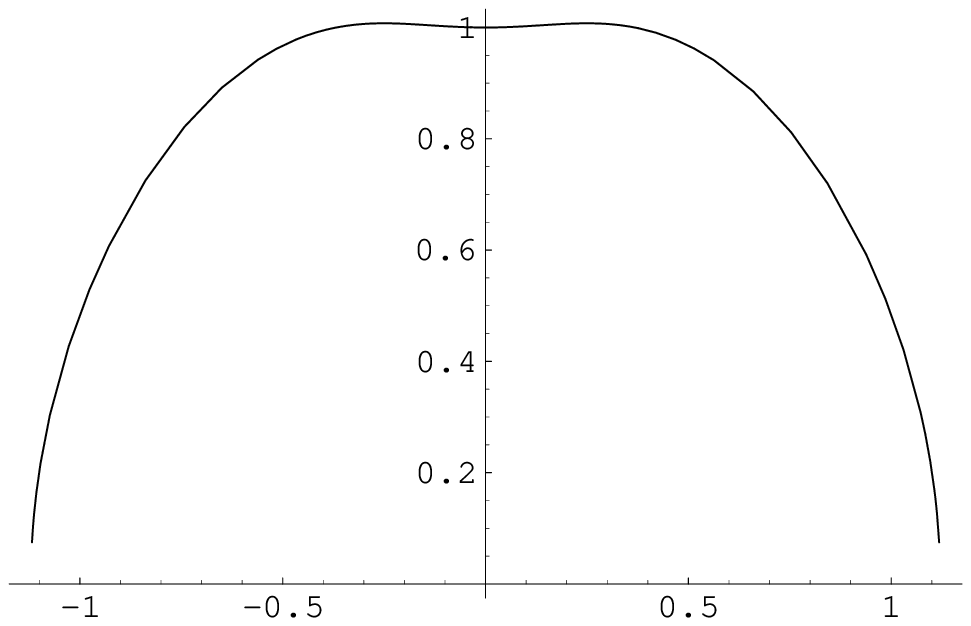,width=3.5cm},
\epsfig{figure=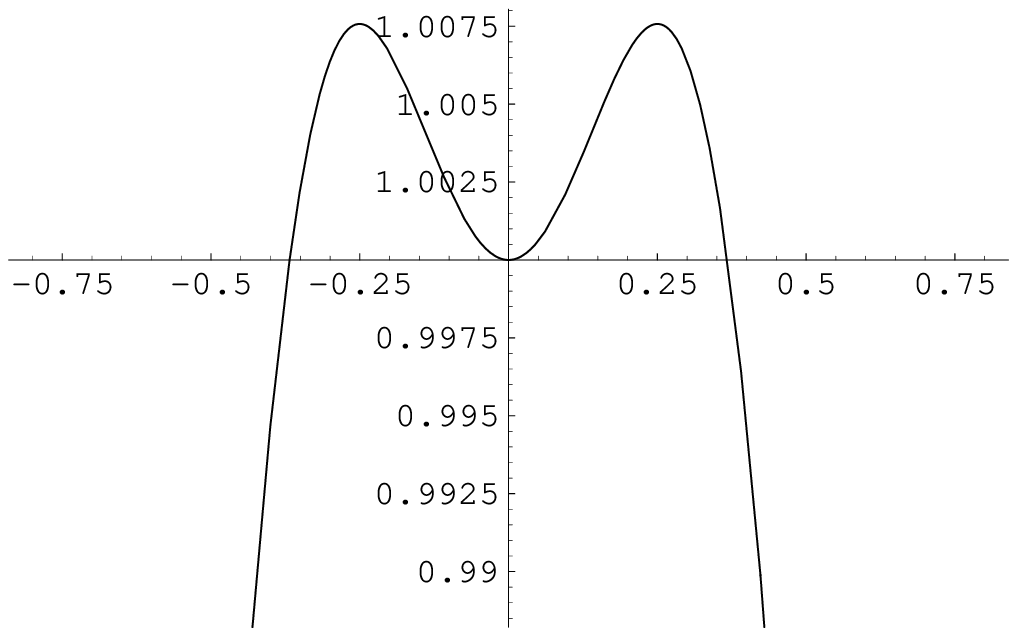,width=3.5cm},
\epsfig{figure=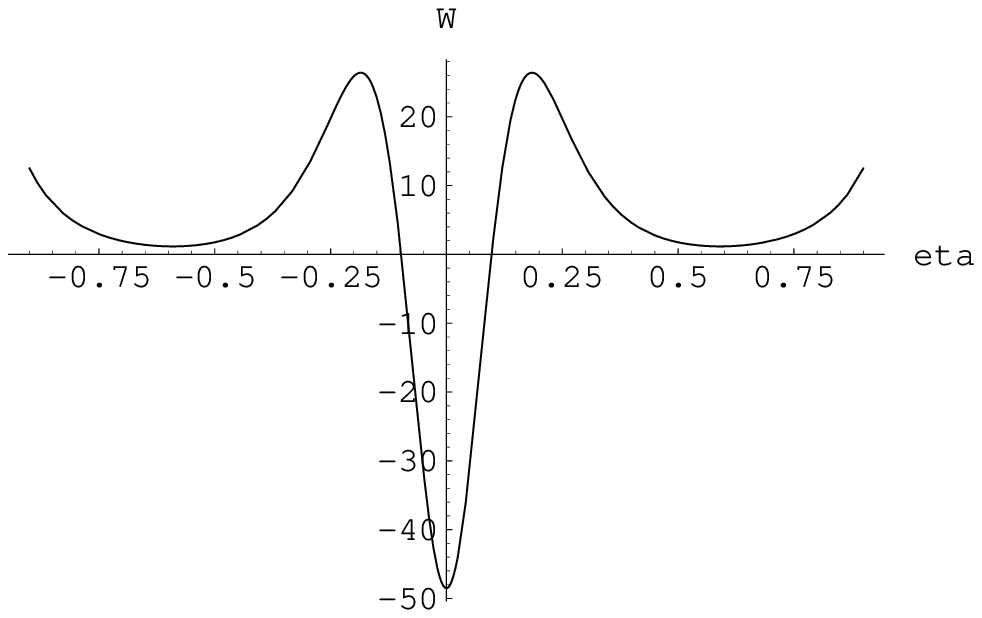,width=4cm}} 
\caption{{\it Lorentzian scalar potentials (2).}  The left panel shows the scalar potential $V(\varphi)={v\over 1+c^2\varphi^2}$ with $v=2.5$ (hence $\Upsilon=0.5$) and $c^2=10$. The central panels show the corresponding scale factor. The right panel shows the
potential for the matter perturbations, $W(\eta)$, which is similar to
Figure 3.}
\label{Figure 4}
\end{figure}

\bigskip
\noindent
{\csc V. Conclusion}
\vskip 0.5cm

In summary, we found that the main result of [1], that is the modification of the spectrum of matter perturbations through a bounce due to a deep potential well, is quite robust,
{\it provided} two hypotheses are fulfilled~: (1), the scalar field conformal gradient remains small at the bounce ($\Upsilon\ll1$) and, (2),  the effective potential for the
perturbations,
$W(\eta)$, decays for $|\eta|\gg\sqrt\Upsilon$, so that asymptotic regions for the in and out states can be defined. 

We also found that these hypotheses,
with the additional requirement that the universe does not recollapse prematurely, yield  a ``conical hat" scalar field potential, with an effective squared
mass  which is negative and large in comparison with the scale set by the radius of the universe at the bounce (but which can be small in Planck units)~; more peculiar is the
fact that the scalar field potential seems to have to decay almost linearly (before tending to zero in the asymptotic region) in order to prevent premature recollapse. 

Finally, we examined hypothesis (1) ($\Upsilon\ll1$) and found that it was not necessary if one only imposed the {\it total} energy density of the scalar field to remain small (in
Planck units) in the whole bouncing region. Now, when all parameters are of order one, the spectrum of the perturbations is either unaffected by the bounce, as stated in [1] or, if
the decay of the scalar field potential $V(\varphi)$ is too fast,  the universe recollapses immediatly and no definite conclusion can be drawn on the modification of the
perturbation spectrum. However a sharply peaked potential $V(\varphi)$, with $2<V(0)<3$ (in units set by the size of the universe at the bounce),  which decays slowly enough,
``does the job" and ensures a modification of the perturbation spectrum.

The question now is~: do there exist plausible (ideally string-inspired) scalar field potentials which possess such properties. In this short Note we shall leave that difficult
problem open.

\vskip 1cm
\noindent
{\bf Acknowledgements}. The author warmly thanks J\'er\^ome Martin and Patrick Peter for discussions on their paper [1], which led to the first version of this Note.
Recent discussions on their criticisms  led me to send them a second version of my Note, which expanded a number of points to try and make them clearer. Unfortunately
 that did not seem to have convinced the authors of [1]~: see their Web Note gr-qc/0406062. It is that second version that is presented here.

\bigskip
\noindent
{\csc Notes}
\vskip 0.5cm

[7] Hence, if, say, $l=3$ denotes a length that we want to compare to the Planck length $l_{Planck}$, then the dimensionful length corresponding to $l=3$ is
$l_{ph}=3\,a_0\, l_{Planck}$~; similarly if $m=3$ denotes a mass, then the dimensionful mass is $m_{ph}=3\,m_{Planck}/a_0$, and,  if $\varrho=3$ denotes an
energy density, then the dimensionful density is $\varrho_{ph}=3\varrho_{Planck}/a_0^2$, etc. Finally, if $\Upsilon$ is a dimensionless number, then $\Upsilon=3$
means... $\Upsilon=3$. Consequently, when in the following we refer to a dimensionful quantity as ``small" or ``large", we shall mean small or large with respect to the
scales set by $a_0$. Now (and I thank Jerome Martin and Patrick Peter for pointing that out to me), if $a_0$ is a large number (that
is if the size of the universe at the bounce is large with respect to the Planck length) the scales set by $a_0$ are small compared to the Planck scales.

[8] In fact, in order for the perturbation equation (2.2) to make sense, $\varphi'$ must be different from zero in the whole bouncing region. The approach developped here
and in [1] therefore does not apply to the problem studied by C. Gordon, N. Turok, Phys. Rev. D {\bf 67} (2003) 123508 (see also N. Deruelle and A. Streich, gr-qc/0405003).

[9] There is of course nothing  a priori ``wrong" in approximating the scale factor by a polynomial. Indeed what happens is the following. Since $\Upsilon$ is supposed to be small,
there are two ranges of variation for $\eta$~: the bouncing region $|\eta|\ll1$ (in practice $|\eta|\leq1$) where the scale factor (which is supposed to be a smooth function of
time) is well approximated by a Taylor expansion in $\eta$ (to order $4$ say), and the corresponding range of validity for the
perturbation potential
$W(\eta)$, which is much smaller,
$|\eta|\leq\sqrt\Upsilon$. In order to obtain $W(\eta)$ outside its range of validity set by the Taylor expansion of the scale factor (to order $4$ say), more information is a
priori required about the scale factor (its Taylor expansion to order $2N\simeq 1/\Upsilon$ say). Truncating the Taylor expansion, that is to say approximating the scale factor
by a ($4$th order) polynomial is just a simple way to provide that missing information and  is a perfectly valid thing to do {\it if} the Taylor series Lagrange remainder, that is
the time derivative of the scale factor of order 5 (say) is small on the {\it whole range} $|\eta|\in [0,1]$.   Moreover it guarantees  that the effective potential for the
perturbations decays at the outskirts of the bouncing region, so that the transition matrix for the perturbation spectrum can be defined. 
(A particular example, studied in [1], of a scale factor which can be approximated by a truncated Taylor series  of order 4 is
$a(\eta)=\sqrt{1+\eta^2}=1+\eta^2/2-\eta^4/8+3\eta^6/16+...$ for which the Lagrange remainder is small in the whole bouncing region.)
See however Section 3 for the implications of such assumptions on the shape of the potential $V(\varphi)$.

[10] If it happens that $A_n^{\rm pre}=-B_n^{\rm pre}$, then the expansion at next order in
$k/\sqrt{V_m}$ yields $A_n^{\rm post}=-B_n^{\rm post}= Const. A_n^{\rm pre}$, where the $Const.$ is of order one. In that case then, the amplitude of the pre-bounce modes is
not amplified and its spectrum is not modified by the bounce.

[11] In Planck units, $V(\eta)$ in (3.2-4) must be replaced by $a_0^2\,V(\eta)$ where $a_0$ is the radius of the universe at the bounce in units of the Planck length, see Note [7].

[12] Choosing to impose the scale factor to approach its quasi-de Sitter expression in the transition region (instead of being a polynomial) yields a potential $V(\varphi)$
similar to Figure 2.

\end{document}